\documentclass[aps,prl,10pt,twocolumn,longbibliography,noeprint,floatfix,superscriptaddress]{revtex4-2}

\usepackage{amsmath, amssymb}
\usepackage{mathtools}
\usepackage{lipsum}
\usepackage{times}
\usepackage{graphicx}
\usepackage{wasysym}
\usepackage{bm}
\usepackage{hyperref}
\usepackage{xspace}
\usepackage{siunitx}
\usepackage[dvipsnames]{xcolor}
\usepackage{soul}
\usepackage{centernot}
\usepackage{braket}
\usepackage{microtype}
\usepackage{upgreek}
\usepackage{booktabs}
\usepackage{float}

\definecolor{customcolor}{rgb}{0.261, 0.212, 0.658}

\definecolor{myColor2}{rgb}{0.02,0.12,0.3}

\definecolor{myColor}{rgb}{0.02,0.12,0.7}
\definecolor{myciteColor}{rgb}{0.39,0.7,0.89}
\hypersetup{colorlinks=true, linkcolor=myColor, filecolor=myColor, urlcolor=myColor, citecolor=myColor, urlcolor=myColor}
\DeclareSIUnit{\gauss}{\ensuremath{\mathrm{G}}}
\DeclareSIUnit{\bohrradius}{\ensuremath{a_0}}
\graphicspath{{./Figures/}}

\DeclareSIUnit{\nK}{\nano\kelvin}
\DeclareSIUnit{\um}{\micro\metre}
\DeclareSIUnit{\aB}{\emph{a}_0}
\DeclareSIUnit{\G}{G}

\newcommand{\appropto}{\mathrel{\vcenter{
  \offinterlineskip\halign{\hfil$##$\cr
    \propto\cr\noalign{\kern2pt}\sim\cr\noalign{\kern-2pt}}}}}

\def\be{\begin{equation}}
\def\ee{\end{equation}}

\makeatletter
\def\@fnsymbol#1{\ensuremath{\ifcase#1\or *\or \dagger\or \ddagger\or
   \mathsection\or \mathparagraph\or \|\or **\or \dagger\dagger
   \or \ddagger\ddagger \else\@ctrerr\fi}}
\makeatother

\newcommand{\um}{\upmu{\rm m}}
\newcommand{\nK}{\textrm{nK}}
\newcommand{\kB}{k_{\textrm{B}}}

\newcommand{\potassium}{^{39}\textrm{K}}

\newcommand{\ba}{\textbf{a}}
\newcommand{\bb}{\textbf{b}}
\newcommand{\bc}{\textbf{c}}
\newcommand{\bd}{\textbf{d}}

\begin{document} 
 
\title{
{Analog} Simulation of Massive Relativistic {Quantum} Fields in 2 + 1 Dimensions
}
\author{Yansheng Zhang} \email{yz661@cam.ac.uk}
\affiliation{Cavendish Laboratory, University of Cambridge, J. J. Thomson Avenue, Cambridge CB3 0US, United Kingdom}
\author{Feiyang Wang}
\affiliation{Cavendish Laboratory, University of Cambridge, J. J. Thomson Avenue, Cambridge CB3 0US, United Kingdom}
\author{Paul H. C. Wong}
\affiliation{Cavendish Laboratory, University of Cambridge, J. J. Thomson Avenue, Cambridge CB3 0US, United Kingdom}
\author{Alexander C. Jenkins}
\affiliation{Kavli Institute for Cosmology, University of Cambridge, Madingley Road, Cambridge CB3 0HA, UK}
\affiliation{DAMTP, University of Cambridge, Wilberforce Road, Cambridge CB3 0WA, UK}
\author{ {Yi Jiang}}
\affiliation{Cavendish Laboratory, University of Cambridge, J. J. Thomson Avenue, Cambridge CB3 0US, United Kingdom}
\author{Konstantinos Konstantinou} 
\affiliation{Cavendish Laboratory, University of Cambridge, J. J. Thomson Avenue, Cambridge CB3 0US, United Kingdom}
\author{ {Gehrig Carlse}} 
\affiliation{Cavendish Laboratory, University of Cambridge, J. J. Thomson Avenue, Cambridge CB3 0US, United Kingdom}
\author{Nishant Dogra} 
\affiliation{Cavendish Laboratory, University of Cambridge, J. J. Thomson Avenue, Cambridge CB3 0US, United Kingdom}
\author{Joseph H. Thywissen} 
\affiliation{Department of Physics and CQIQC, University of Toronto, Toronto, Ontario M5S 1A7, Canada}
\author{Christoph Eigen}
\affiliation{Cavendish Laboratory, University of Cambridge, J. J. Thomson Avenue, Cambridge CB3 0US, United Kingdom}
\author{Zoran Hadzibabic} \email{zh10001@cam.ac.uk}
\affiliation{Cavendish Laboratory, University of Cambridge, J. J. Thomson Avenue, Cambridge CB3 0US, United Kingdom}

\begin{abstract}
Quantum field theories provide fundamental models of complex interacting systems~\cite{Peskin:1995, Wen:2004, Weinberg:2008, Altland:2010}, from high-energy physics and cosmology to condensed matter. 
However, solving these models in non-perturbative and dynamical regimes is often challenging, particularly in more than one spatial dimension.
Analog simulation using tunable synthetic quantum systems can both verify existing theoretical predictions and lead to new physical insights.
Here, we realize { analog} simulation of massive relativistic { quantum} fields in $2+1$ dimensions (two spatial dimensions and time), using two coherently coupled spin components~\cite{Raghavan:1999, Recati:2022} in a uniform two-dimensional atomic Bose--Einstein condensate~\cite{Chomaz:2015,Navon:2021}. Specifically, we encode the paradigmatic sine-Gordon model~\cite{Weinberg:2012, Samaj:2013} in the { long-wavelength modes of the} field describing the relative phase, $\phi$, of the two components. In the perturbative regime, collective field excitations exhibit a relativistic dispersion with a tunable mass gap. We also observe explicitly non-perturbative phenomena, including the existence of topological domain walls across which $\phi$ rapidly winds by $2\pi$~\cite{Son:2002}. Our work opens possibilities for studies of cosmologically relevant phenomena including preheating~\cite{Kofman:1994, Chatrchyan:2021}, dynamics of topological defects~\cite{Vilenkin:1985, Vilenkin:2000}, and relativistic false-vacuum decay~\cite{Coleman:1977a, Callan:1977, Fialko:2015, Braden:2018, Billam:2019,Braden:2019, Jenkins:2024}.
\end{abstract}
\maketitle  

Analog simulation offers insights into quantum field theories (QFTs) pertaining to phenomena in naturally inaccessible settings, such as the early universe.
Time evolution in a QFT simulator is especially valuable due to the computational complexity of non-perturbative many-body dynamics.
Single-component atomic Bose--Einstein condensates (BECs), with tunable microscopic Hamiltonians, provide a simulation platform for massless relativistic fields and have been used to simulate cosmological acoustic fluctuations~\cite{Jaskula:2012, Hung:2013, Clark:2017, Eckel:2018, Viermann:2022, Gondret:2025, Schutzhold:2025}.
Extending experiments to two-component BECs with coherent inter-component coupling~\cite{Raghavan:1999, Albiez:2005, Levy:2007, Zibold:2010,  Trenkwalder:2016, Spagnolli:2017,Recati:2022, Cominotti:2024}, one can 
simulate QFTs with massive fields and compact conjugate variables, which can display long-wavelength non-perturbative dynamics~\cite{Schumm:2005, Betz:2011, Nicklas:2015, Schweigler:2017, Pignuer:2018, Farolfi:2021a, Ji:2022, Cominotti:2022, Cominotti:2023, TiZhang:2024, Zenesini:2024, Cominotti:2025, Rydow:2025} and support broader families of topological defects~\cite{Son:2002,Vachaspati:2006, Weinberg:2012}. In particular, this allows realization of the paradigmatic sine-Gordon model~\cite{Weinberg:2012} for massive relativistic fields. Previously, this was achieved in $d=1$ dimension~\cite{Schumm:2005, Betz:2011, Schweigler:2017, Pignuer:2018, Ji:2022, TiZhang:2024}, where the model (for a uniform system) is integrable~\cite{Samaj:2013}. 
Here, we realize quantum simulation of massive relativistic fields in $d=2$ and observe both perturbative and non-perturbative sine-Gordon dynamics.

\begin{figure}[!h]
    \centering
    \includegraphics[width=\columnwidth]{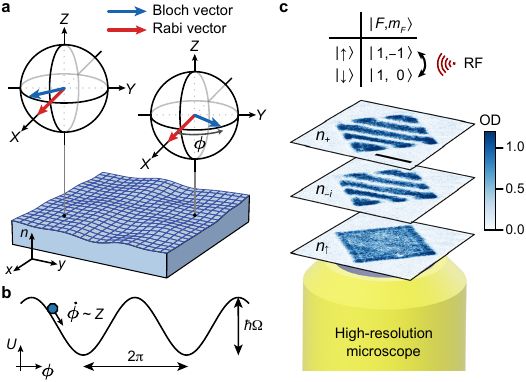}
    \caption{\textbf{Two-dimensional simulator for massive relativistic fields.} 
     \ba, The concept. In our two-dimensional BEC, the density $n$ is essentially uniform, and we use a spatially varying spin state, shown by the Bloch vectors, to encode a massive relativistic field. The conjugate variables $Z(x,y)$ and $\phi(x,y)$ are, respectively, the local population imbalance and relative phase of spin components $\ket{\uparrow}$ and $\ket{\downarrow}$, which are coherently coupled by a radio-frequency (RF) field, indicated by the Rabi vector. 
     \bb, Sine-Gordon model.
     In the Josephson regime, $\mu_\text{s} \gg \hbar \Omega$, where $\mu_\text{s}$ is the spin chemical potential and $\Omega$ the RF Rabi frequency, $\phi$ evolves under a sinusoidal potential $U$.
    \bc, The experiment. 
      Our $\ket{\uparrow}$ and $\ket{\downarrow}$ are two hyperfine states of $^{39}$K. 
    We image both the local populations $n_{\uparrow,\downarrow} = |\psi_{\uparrow,\downarrow}|^2$ (where $\psi_{\uparrow,\downarrow}$ are the wavefunctions) and the transverse spin projections $n_{+,-}=|\psi_\uparrow\pm \psi_\downarrow|^2/2$ and $n_{i,-i}= |{ i}\psi_\uparrow\pm \psi_\downarrow|^2/2$. Here, we show example images of $n_+$, $n_{ -i}$, and $n_\uparrow$, for a cloud with {  uniform $n$ and $Z$ but a magnetically imprinted linear $\phi$ gradient}.
    The scale bar corresponds to $20~\upmu$m; OD is optical density.
    }
    \label{fig:1}
\end{figure}

Our simulator is based on a coherently coupled mixture of two miscible hyperfine states of $\potassium$ atoms in a two-dimensional optical box trap~\cite{Chomaz:2015, Navon:2021, Christodoulou:2021}; {  we denote the two states $\ket{\uparrow}$ and $\ket{\downarrow}$, and their individual wavefunctions $\psi_{\uparrow,\downarrow}=\sqrt{n_{\uparrow,\downarrow}}e^{-i\phi_{\uparrow,\downarrow}}$, where $n_{\uparrow,\downarrow}$ and $\phi_{\uparrow,\downarrow}$ are densities and phases of the individual components. In our gas, the two-dimensional total density $n=n_{\uparrow}+n_{\downarrow}$ is essentially uniform, but the spin state varies in space (Fig.~\ref{fig:1}{\ba}); on the Bloch sphere, the two compact conjugate variables are the local population imbalance $Z(x,y)=[n_{\uparrow}(x,y)-n_{\downarrow}(x,y)]/[n_{\uparrow}(x,y)+n_{\downarrow}(x,y)]$ and the relative phase of the two components $\phi(x,y)=\phi_{\uparrow}(x,y)-\phi_{\downarrow}(x,y)$.}
The strength of the coherent coupling is given by the Rabi frequency $\Omega$, while nonlinear interactions set the spin chemical potential $\mu_{\rm s}$ { (Methods)}. In the Josephson regime, $\mu_{\rm s} \gg \hbar\Omega$ (where $\hbar$ is the reduced Planck constant), fluctuations in $Z$ are strongly suppressed, and long-wavelength spin dynamics are dominated by fluctuations in $\phi$. At constant $Z$, the energy cost for $\phi$ to deviate from $0$ (defined by the Rabi vector) corresponds to a potential
$U(\phi) = - (\hbar \Omega/2) \cos\phi$ (Fig.~\ref{fig:1}{\bb}), which leads~\cite{Jenkins:2024} to the sine-Gordon equation, 
\begin{equation}
    \partial^2_t\phi =  c^2\,\nabla^2\phi -  ( m^{*}c^2/\hbar )^2~\sin\phi \, ,
    \label{eq:SG model}
\end{equation}
with the spin sound speed $c = \sqrt{\mu_{\rm s}/m}$, where $m$ is the atom mass, and the field mass $m^* = m \sqrt{2\hbar\Omega/\mu_{\rm s}}$, which arises from the coherent coupling that breaks the $U(1)$ symmetry of $\phi$.
Solutions of this model include domain walls where $\phi$ rapidly winds by $2\pi$~\cite{Son:2002,Vachaspati:2006, Weinberg:2012}. While in $d=1$ such walls are just point defects, in $d>1$ they are extended objects that can exhibit nontrivial patterns~\cite{Vilenkin:1985, Vilenkin:2000}.

\begin{figure}
    \centering
    \includegraphics[width=\columnwidth]{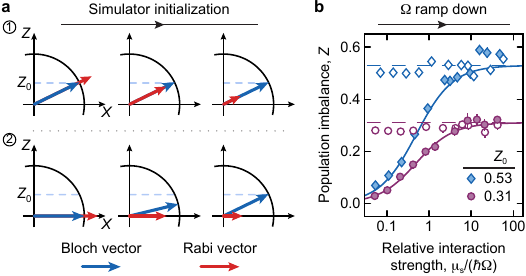}
    \caption{\textbf{Robust simulator initialization.}
    \ba, Experimental protocols.
    Spin and total-density dynamics decouple for $Z=Z_0$ (see text). In  protocol {\textcircled{\raisebox{-.9pt} {1}}}, we prepare the system at the predicted $Z_0$ using strong RF coupling, and then reduce this coupling to reach the Josephson regime, $\mu_{\rm s} \gg\hbar \Omega$. Alternatively, in {\textcircled{\raisebox{-.9pt} {2}}}, we start with a strongly coupled symmetric mixture ($Z=0$) and, as we reduce $\Omega$, a nonzero $Z$ spontaneously emerges and tends towards $Z_0$.
    \bb,  Experimental data for protocols {\textcircled{\raisebox{-.9pt} {1}}} (open symbols) and {\textcircled{\raisebox{-.9pt} {2}}} (solid symbols), for two sets of interaction parameters corresponding to different $Z_0$ values (Methods). 
    The dashed lines show the predicted $Z_0$, and the solid lines show predictions based on {  $\chi={{\rm d}Z}/{{\rm d}\delta}\simeq\left[{\Omega + 2\mu_{\rm s}/\hbar}\right]^{-1}$, where $\delta$ is the RF detuning}; for the same $\Omega$, the RF field in protocol {\textcircled{\raisebox{-.9pt} {2}}} is detuned from that in {\textcircled{\raisebox{-.9pt} {1}}} by $-Z_0\Omega$, so $Z \simeq Z_0 (1 -  \chi \Omega)$.
    { 
    Each data point is based on $10$ independent experimental runs; the symbols show the mean values and the error bars (often smaller than symbol sizes) denote the s.e.m.
    }
    }
    \label{fig:2}
\end{figure}

In our experimental setup (Methods), $\ket{\uparrow} = \ket{1, -1}$ and $\ket{\downarrow} = \ket{1, 0}$ in the $\ket{F, m_F}$ basis, and we couple them by a radio-frequency (RF) field with $\Omega$ up to $2\pi\times 20\,$kHz. With our $B$ field stability of $\approx 150\,\upmu$G, our error in the RF detuning $\delta$ is $\delta_\text{err}\simeq 2\pi \times 100\,$Hz.
The contact interactions between atoms in $\sigma,\sigma'\in\{\uparrow,\downarrow\}$ are characterized by parameters $g_{\sigma\sigma'}\propto a_{\sigma\sigma'}$, where $a_{\sigma\sigma'}$ are $s$-wave scattering lengths. We operate near $B = 58\,$G, where $a_{\uparrow\uparrow}\simeq32\,a_0$ and $a_{\uparrow\downarrow}\simeq-53\,a_0$ (where $a_0$ is the Bohr radius) are essentially fixed, and $a_{\downarrow\downarrow}$ is tunable via a Feshbach resonance. We tune $a_{\downarrow\downarrow} > 85\,a_0$ to ensure that $a_{\uparrow\uparrow}a_{\downarrow\downarrow}>a_{\uparrow\downarrow}^2$, so that the mixture is miscible and stable against mean-field collapse~\cite{Cabrera:2018}. In our experiments, $n \sim 10^2~\upmu$m$^{-2}$ and $\mu_{\rm s}=\kappa n\simeq 2\pi \hbar \times 800\,$Hz, where $\kappa=(g_{\uparrow\uparrow}+g_{\downarrow\downarrow}-2g_{\uparrow\downarrow})/4$. 
{  Note that we start our experiments with a quasi-pure condensate; from our trap depth of $\approx \kB \times 30\,$nK (where $\kB$ is the Boltzmann constant), we estimate the temperature to be $\lesssim 10\,$nK, while the Berezinskii--Kosterlitz--Thouless critical temperature corresponding to our $n$ is $T_{\rm BKT} \approx 0.5\,\upmu$K~\cite{Hadzibabic:2011}.}

Using in-situ state-selective absorption imaging (Fig.~\ref{fig:1}\bc), we directly image $n_{\uparrow,\downarrow} (x,y) = |\psi_{\uparrow,\downarrow}(x,y)|^2$. By applying suitable $\pi/2$ spin rotations~\cite{Sadler:2006}, just prior to the imaging, we also measure $n_{+,-}(x,y)=|\psi_\uparrow\pm \psi_\downarrow|^2/2$ and $n_{i,-i}(x,y) = | i\psi_\uparrow\pm \psi_\downarrow|^2/2$. 
Additionally, to measure projections of the global Bloch vector ($X$, $Y$, or $Z$), averaged over the system, we (after any necessary $\pi/2$ rotations) release the atoms from the trap, spatially separate $\ket{\uparrow}$ and $\ket{\downarrow}$ components with a Stern--Gerlach gradient, and image them simultaneously.

For general $g_{\sigma\sigma'}$, spin-dependent interaction energy {  is minimized for $Z = Z_0$, where $Z_0 = -\Delta/\kappa$, with $\Delta=(g_{\uparrow\uparrow}-g_{\downarrow\downarrow})/4$ (Methods)}. For $Z=Z_0$, the interaction contribution to the chemical potentials of the two components, $\mu_\uparrow$ and $\mu_\downarrow$, is equal. Consequently, total-density fluctuations do not change $\mu_\uparrow - \mu_\downarrow$, and thus (to leading order) do not couple to the spin dynamics~\cite{Jenkins:2024}, which is essential for realizing the sine-Gordon model in the spin sector. Note that $Z_0 = 0$ for a $\mathbb{Z}_2$-symmetric system ($\Delta = 0$), while in our case $\Delta < 0$, so $Z_0 > 0$.   

\begin{figure}
    \centering
    \includegraphics[width=\columnwidth]{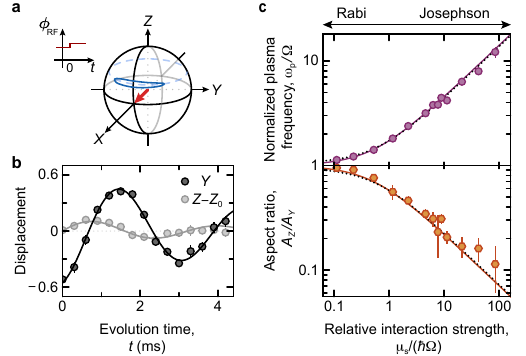}
    \caption{\textbf{Plasma oscillations.} \ba, A jump in the RF-field phase, $\phi_{\rm RF}$, initiates global spin oscillations. Interactions squash the oscillation trajectory (blue ellipse) in the $Z$ direction. \bb, An example of $Y$ and $Z$ oscillations, for  $\mu_{\rm s}/(\hbar\Omega)\simeq11$; note that $\phi \simeq Y$. Damped-sine fits (solid lines) give the plasma frequency $\omega_\text{p}$ and the oscillation amplitudes $A_Y$ and $A_Z < A_Y$.
    { 
    Each data point is based on $5$ independent experimental runs; the symbols show the mean values and the error bars (often smaller than the symbol sizes) denote the s.e.m.}
    \bc, Dependence of $\omega_\text{p}/\Omega$ and $A_Z/A_Y$ on $\mu_\text{s}/(\hbar\Omega)$. The Rabi and Josephson regimes, respectively, correspond to $\mu_\text{s}/(\hbar\Omega) \ll 1$ and $\gg 1$. 
    For the measurements here, $Z_0=0.31$,
    { 
    and the dotted lines show the corresponding theoretical predictions (Methods), while the essentially indistinguishable solid lines show the predictions for $Z_0=0$ from Eqs.~(\ref{eq:aspect-ratio}) and (\ref{eq:plamsa-freq}).}
     { 
    The error bars show fitting errors.}
    }
    \label{fig:3}
\end{figure}

\begin{figure*}
    \centering
    \includegraphics[width=\textwidth]{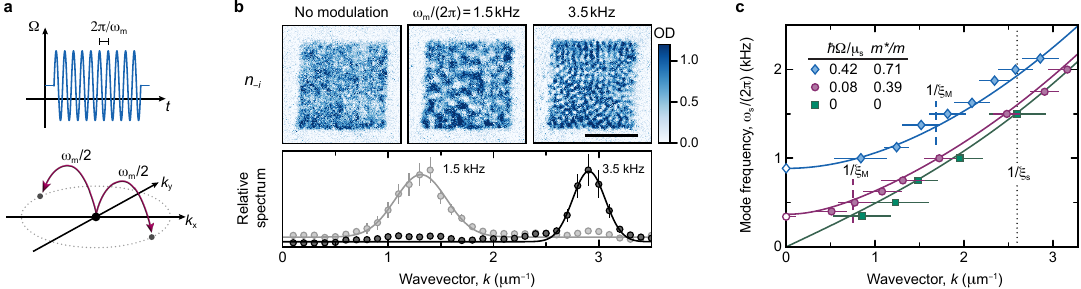}
    \caption{\textbf{Massive relativistic dispersion relation.} \ba, Experimental protocol. 
    We parametrically excite the system by modulating $\Omega$ at frequency $\omega_\text{m}$. The injected energy, in quanta of $\hbar\omega_\text{m}$, is converted to pairs of spatial excitations with opposite momenta, each carrying energy $\hbar\omega_\text{m}/2$. 
    \bb, Example measurements. Top: After the parametric modulation, images of $n_{-i}$ (top) show fluctuations with a characteristic $\omega_\text{m}$-dependent length scale; the scale bar corresponds to $20\,\upmu$m.
    Bottom: Normalizing the spatial power spectrum of the fluctuations observed with modulation to that without modulation, we reveal the dominant $k$ excited at each $\omega_{\rm m}$, which gives $\omega_{\rm s}(k) = \omega_{\rm m}/2$. {  Each data point is based on $5$ independent experimental runs; the symbols show the mean values and the error bars denote the s.e.m.} The solid lines show Gaussian fits.
    \bc, Dispersion relations for different field masses $m^*$, tuned by varying the value of $\Omega$ (around which we modulate); here we also include $k=0$ plasma-oscillation measurements.  
    The solid lines show theoretical predictions from Eq.~(\ref{eq:dispersion}). The dashed lines indicate the inverse magnetic healing lengths $1/\xi_\text{M}=\sqrt{2m\Omega/\hbar}$; in our system, $\xi_\text{M}$ corresponds to the reduced Compton wavelength $\hbar/(m^*c)$. The dotted line shows the inverse spin healing length, $1/\xi_{\rm s} = \sqrt{2m\mu_\text{s}/\hbar^2}$, {  which sets the low-$k$ range where the dispersion is approximately relativistic (see text)}.
    The error bars in $k$ denote the Gaussian widths of the spectra such as shown in {\bb}.
    For these measurements $Z_0=0.31$.
    }
    \label{fig:4}
\end{figure*}

In Fig.~\ref{fig:2}, we show how we initialize our simulator at $Z_0$, for different $Z_0$ values. In one scenario, we first use strong RF coupling to adiabatically rotate the Bloch vector from state $\ket{\uparrow}$ to the theoretically predicted $Z_0$. Then, as we slowly (over $10\,$ms) reduce the RF coupling to reach the Josephson regime, $Z$ remains constant; see \raisebox{.5pt}{\textcircled{\raisebox{-.9pt} {1}}} in Fig.~\ref{fig:2}{\ba} and open symbols in Fig.~\ref{fig:2}{\bb}. Note that here we proportionally reduce $\Omega$ and $\delta$, so that the Rabi vector $(\Omega, 0, \delta)$ remains aligned to $Z_0$. Alternatively, if we instead prepare a strongly coupled symmetric mixture ($Z=\delta = 0$) and then slowly reduce $\Omega$, the interactions `torque' the Bloch vector towards $Z_0$; see \raisebox{.5pt}{\textcircled{\raisebox{-.9pt} {2}}} in Fig.~\ref{fig:2}{\ba} and solid symbols in Fig.~\ref{fig:2}{\bb}. Remarkably, one does not need to know the correct $Z_0$ {\it a priori} -- the system spontaneously and robustly finds it as the ground state in the Josephson regime.

This robustness also highlights an experimentally important reduced sensitivity to $\delta_\text{err}$. In a non-interacting system, spin control with error $Z_\text{err} \ll 1$ requires $\delta_\text{err} \ll \Omega$, which can pose challenging demands for the $B$-field stability~\cite{Cominotti:2024}. However, here, interactions reduce the magnetic susceptibility~{ $\chi={{\rm d}Z}/{{\rm d}\delta}\simeq\left[{\Omega + 2\mu_{\rm s}/\hbar}\right]^{-1}$}~\cite{Farolfi:2021b,Recati:2022}, so $Z$ is robust as long as  $\delta_\text{err} \ll 2\mu_{\rm s}/\hbar$, and we achieve $Z_\text{err} <0.07$ { (r.m.s.)} even for $\Omega \ll \delta_\text{err}$.

In Fig.~\ref{fig:3} we examine the global spin dynamics in our system, for $Z_0=0.31$. By jumping the phase of the RF field (which defines $\phi=0$), we globally displace the Bloch vector and excite plasma oscillations (Fig.~\ref{fig:3}\ba), such as seen in Fig.~\ref{fig:3}\bb.

\begin{figure*}
    \centering
    \includegraphics[width=\textwidth]
    {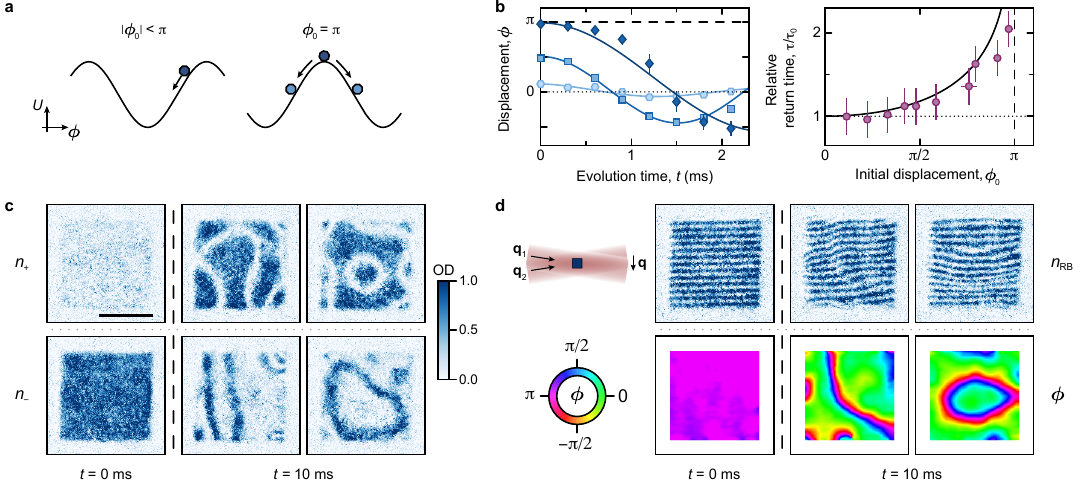}
    \caption{
    \textbf{Non-perturbative sine-Gordon dynamics.} \ba, Experimental concepts. We study non-perturbative dynamics following a large initial phase displacement $\phi_0$. Left: For large $|\phi_0|<\pi$, the concavity of the sine-Gordon potential $U(\phi)$ leads to an amplitude-dependent oscillation period. Right: for $\phi_0=\pi$, different regions of the system may eventually settle to different potential minima, which results in formation of domain walls across which $\phi$ rapidly winds by $2\pi$.
    \bb, Large-amplitude plasma oscillations.  Left: We measure $\phi(t)$ for different $|\phi_0| < \pi$ and extract the time, $\tau$, for the system to return to $\phi =0$. 
    Solid lines show sinusoidal fits. Note that for the largest $\phi_0$ here the oscillation is visibly anharmonic. {  Each data point is based on $5$ independent experimental runs, and the error bars show the s.e.m.}
    Right: $\tau/\tau_0$ versus $\phi_0$; here $\tau_0 = \pi/(2\omega_{\rm p})$ is the $\phi_0$-independent result in the perturbative (harmonic) regime. The solid line is the theoretical prediction based on the mechanical-pendulum model. {  The error bars in $\tau$ correspond to half of the spacing between the sampled $t$ values.} For these measurements, $Z_0 = 0.31$, $\Omega/(2\pi) \simeq 74\,$Hz,  and $\mu_\text{s}/(\hbar\Omega)\simeq 11$. 
    {\bc} and {\bd}, The domain walls. The system is prepared at $\phi\simeq\pi$ at $t=0$, and allowed to relax for $10\,$ms, corresponding to $\approx 7\tau_0$; here $Z_0 = 0.53$, $\Omega/(2\pi) \simeq 22\,$Hz, and $\mu_{\rm s}/(\hbar\Omega)\simeq39$. The black scale bar shows $20\,\upmu$m.  In \bc, we show examples of domain walls seen in the $n_+$ and $n_-$ images, where they appear as, respectively, dark and white lines; each image shows an independent realization {  (see also Extended Data Fig.~\ref{fig:ExtendedFigure3} for complementary images taken in the same experimental run)}.  
    {  In \bd, we show phase maps that explicitly reveal the $2\pi$ phase winding at domain walls. 
    We interfere the two spin states using a pair of Raman-Bragg laser beams with wave vectors $\mathbf{q}_{1,2}$, and the imaged density distribution, $n_\text{RB}$, shows modulations $\propto \cos(\mathbf{q}\cdot\mathbf{r} + \phi)$, where $\mathbf{q} = \mathbf{q}_1 - \mathbf{q}_2$. By analyzing such fringe patterns, we reconstruct the phase maps shown in the bottom panels (see text and Extended Data Fig.~\ref{fig:ExtendedFigure4} for more details).}
    }
    \label{fig:5}
\end{figure*}

Here, neglecting small corrections 
{  due to $Z_0$ not being zero (see Methods)}, the displacements $\delta Z$ and  $\delta Y \simeq \sin \delta\phi \simeq \delta\phi$ evolve under the effective Lagrangian density
\begin{equation}
    \mathcal{L}/n = \frac{1}{2}\hbar\,\delta\dot{\phi}~\delta Z - \frac{1}{4}\hbar\Omega~(\delta\phi)^2- \frac{1}{4}(\hbar\Omega + 2\mu_{\rm s})~(\delta Z)^2  \, .
    \label{eq:langrangian}
\end{equation}
The energy cost of a $\phi$ displacement is only due to $U \propto \hbar\Omega$, but the cost of a $Z$ displacement also includes an interaction-energy penalty, $\propto \mu_{\rm s}$. The ratio of these costs gives the ratio of the oscillation amplitudes:
\begin{equation}
     {A_Z}/{A_\phi} \simeq {A_Z}/{A_Y}\simeq 1/\sqrt{1 + 2\mu_{\rm s}/(\hbar\Omega) } \, ,
     \label{eq:aspect-ratio}
\end{equation}
while their product gives the plasma oscillation frequency
\begin{equation}
\omega_{\text{p}}\simeq \Omega \sqrt{1 + 2\mu_{\rm s}/(\hbar\Omega)} \, .
\label{eq:plamsa-freq}
\end{equation}
In the non-interacting (Rabi) limit ${A_Z} = {A_Y}$ and $\omega_{\text{p}} = \Omega$, while in the Josephson regime ${A_Z}\ll {A_Y}$, which is essential for the implementation of the sine-Gordon model, and $\hbar \omega_{\text{p}}\simeq\sqrt{2\hbar \Omega \mu_{\rm s}} \gg \hbar \Omega$ is the mass gap $m^*c^2$ in Eq.~(\ref{eq:SG model}).

Our measurements (Fig.~\ref{fig:3}\bc) agree with the predictions of Eqs.~(\ref{eq:aspect-ratio}) and (\ref{eq:plamsa-freq}) over three orders of magnitude in $\mu_{\rm s}/(\hbar\Omega)$.
Crucially, we observe coherent oscillations deep in the Josephson regime, up to $\mu_\text{s}/(\hbar\Omega) \sim 10^2$.

Next, we turn to the finite-momentum ($k>0$) spin excitations (Fig.~\ref{fig:4}). By modulating $\Omega$ at frequency $\omega_\text{m}$, we parametrically excite finite-$k$ spin modes with frequency $\omega_\text{s}(k) = \omega_\text{m}/2$ (Fig.~\ref{fig:4}\ba). Note that the coupling of this parametric drive to the total-density modes is much weaker. 

Figure~\ref{fig:4}{\bb} shows example images with and without parametric modulation. We measure $n_{-i}(x,y)$, which reveals the $\phi$ excitations. For every $\omega_\text{m}$ we observe $\phi$ modulations with a well-defined spatial frequency $k$, and thus map out the spin dispersion relation, shown in Fig.~\ref{fig:4}{\bc} for different $m^*$ values. 

The measured dispersion is in good agreement with the theoretical prediction~\cite{Jenkins:2024,Recati:2022}~{  (see also Methods)}
\begin{equation}
    \hbar\omega_{\rm s}\simeq \sqrt{(\epsilon_k+\hbar\Omega)(\epsilon_k+\hbar\Omega + 2\mu_\text{s})} \, ,
    \label{eq:dispersion}
\end{equation}
where $\epsilon_k=\hbar^2k^2/(2m)$. For $k \lesssim 1/\xi_{\rm s} \simeq 2.6~\upmu$m$^{-1}$, where $\xi_\text{s} = \sqrt{\hbar^2/(2m\mu_\text{s}})$ is the spin healing length, this prediction is close to the massive relativistic dispersion relation
\begin{equation}
    \hbar \omega_{\rm r} \simeq \sqrt{\left(m^*c^2\right)^2 + \left(\hbar ck\right)^2} \, ,
    \label{eq:relativistic dispersion}
\end{equation}
{ 
where we now use the more general expressions for $c \simeq \sqrt{(\hbar\Omega + \mu_{\rm s})/m}$ and $m^* \simeq m \sqrt{\hbar\Omega(\hbar\Omega + 2\mu_{\rm s})/(\hbar\Omega + \mu_{\rm s})^2}$, which are  valid for any $\mu_{\rm s}/(\hbar\Omega)$; see Extended Data Fig.~\ref{fig:ExtendedFigure1} for a comparison between Eqs.~(\ref{eq:dispersion}) and (\ref{eq:relativistic dispersion}).
}

For $\Omega=0$, the dispersion is gapless, while a nonzero $\Omega$ opens up a gap at $k=0$ {  (see also Ref.~\cite{Cominotti:2022})}.
For a massive relativistic field, the transition from the low-energy quadratic dispersion to the higher-energy linear dispersion is set by the reduced Compton wavelength $\lambdabar =  \hbar/(m^*c)$, which here corresponds to the magnetic (Rabi) healing length $ \xi_\text{M}=\sqrt{\hbar/(2m\Omega)}$.
Our ability to tune $\mu_{\rm s}/(\hbar\Omega) = \left(\xi_{\rm M}/\xi_{\rm s}\right)^2\gg 1$ gives us a $k$ range where the relativistic nature of the massive dispersion is important.

We now turn to exploring the non-perturbative sine-Gordon dynamics, by initializing the system at a large global phase displacement $\phi_0$ (Fig.~\ref{fig:5}\ba). 

For $|\phi_0| <\pi$, we observe the anharmonic nature of the sine-Gordon potential, $U(\phi) \sim \cos(\phi)$, by measuring the time, $\tau$, it takes the system to return from $\phi_0$ to $\phi=0$ (Fig.~\ref{fig:5}\bb).
Here, we extract $\phi = \tan^{-1}(Y/X)$, where $Y$ and $X$ are the separately measured projections of the global Bloch vector.
In the perturbative regime, $U(\phi)$ is approximately harmonic and $\tau \simeq \pi/(2\omega_\text{p})$ is independent of $\phi_0$. However, as $\phi_0 \rightarrow \pi$, the concavity of $U$ becomes increasingly relevant and $\tau$ grows, in analogy with a mechanical pendulum with a large initial angular displacement (solid line in the right panel of Fig.~\ref{fig:5}\bb). {  Note that at longer times these oscillations damp out, and this damping is accompanied by an emergence of spatial structures corresponding to $k>0$ excitations; see Extended Data Fig.~\ref{fig:ExtendedFigure2}.}

\begin{figure*}
    \centering
    \includegraphics[width=\textwidth]
    {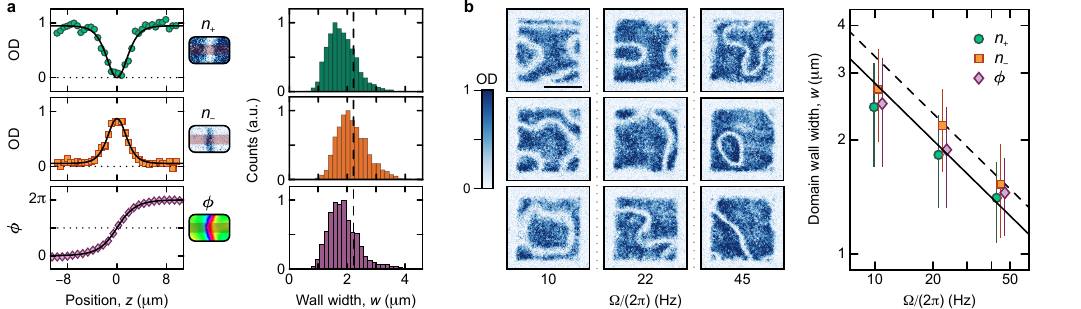}
    \caption{{ 
    \textbf{Domain-wall shape and width.}
    \ba, For the experimental protocol as in Fig.~\ref{fig:5}, we extract the wall widths, $w$, from $n_+$ and $n_-$ images, such as shown in Fig.~\ref{fig:5}{\bc}, and the phase maps, such as shown in Fig.~\ref{fig:5}{\bd}. We fit cuts perpendicular to the walls to the theoretically expected profiles (solid lines, see text). The histograms show width distributions based on $>100$ experimental images, and the dashed lines indicate the theoretically predicted $w$. 
    {\bb}, The effect of changing $\Omega$ on the domain-wall width. Here the experimental protocol is the same, except that for the last $2\,$ms of relaxation we jump $\Omega/(2\pi)$ from $22\,$Hz to either $10\,$Hz or $45\,$Hz. The representative $n_+$ images show an increase of the wall widths in the former case and decrease in the latter; the black scale bar shows $20\,\upmu$m. The summary of the extracted mean widths is consistent with the expected trend $w \propto 1/\sqrt{\Omega}$. 
    The dashed line shows the theoretically predicted $w$, including an $8\%$ correction due to our nonzero $Z_0$ (Methods).
    The solid line shows the same trend with systematically $15\%$ lower values.
    The error bars show the standard deviations of the $w$ distributions. The different symbols corresponding to the same $\Omega$ are slightly offset horizontally for clarity. 
    }}
    \label{fig:6}
\end{figure*}

For $\phi_0=\pi$, different regions of the system may settle into different potential minima, leading to the formation of domain walls across which $\phi$ rapidly winds by $2\pi$. We observe such walls in most experimental realizations, with their spatial patterns varying randomly shot to shot (Fig.~\ref{fig:5}\bc, \bd).

In the images of $n_+$ and $n_-$ (Fig.~\ref{fig:5}\bc), the walls appear as dark and white lines, respectively. {  Here we show images from different experimental realizations, but note that if we image both $n_+$ and $n_-$ in the same experimental run, we see that they are complementary; see Extended Data Fig.~\ref{fig:ExtendedFigure3}.

In Fig.~\ref{fig:5}\bd, we explicitly demonstrate the $2\pi$ phase winding across the domain walls. Here, just prior to imaging the $\ket{\uparrow}$ component, we perform a fast ($\simeq6$\,$\upmu$s) $\pi/2$ spin rotation using a resonant two-photon Raman-Bragg (RB) coupling (see cartoon) instead of an RF field. Compared to the RF field that has a spatially uniform phase across the cloud,  the RB coupling  leads to a complex $\Omega$ with a spatially varying phase $\mathbf{q}\cdot\mathbf{r}$; here $\mathbf{r} = (x,y)$ and $\mathbf{q} = \mathbf{q}_1 - \mathbf{q}_2$, where $\mathbf{q}_{1,2}$ are the wave vectors of the two laser beams. In the Bloch-sphere picture, the  RB Rabi vector always lies in the equatorial plane, but its orientation varies along the direction defined by $\mathbf{q}$. Consequently, the imaged density $n_\text{RB}(x,y)$ shows modulations $\propto \cos(\mathbf{q}\cdot\mathbf{r} + \phi)$. For a spatially uniform $\phi$, this is a pattern of straight parallel fringes with period $2\pi/q\simeq 2.51\,\upmu$m, while spatial variations of $\phi(x,y)$ lead to fringe deformations. In particular, a domain wall perpendicular to $\mathbf{q}$ changes the fringe spacing, while a wall parallel to $\mathbf{q}$ shifts the fringe phase by $2\pi$.
In Fig.~\ref{fig:5}\bd, we show both straight fringes observed at $t=0$ and two examples of domain walls observed after $10\,$ms of relaxation; the bottom panels show the reconstructed phase maps (see Extended Data Fig.~\ref{fig:ExtendedFigure4} for more details).
}

{ 
The characteristic width, $w$, of an equilibrium domain wall is set by the competition between the potential energy per unit length, $\propto \hbar\Omega\cdot nw$, and the corresponding kinetic energy, $\propto\hbar^2/(2mw^2) \cdot nw$. Similarly to the oscillator length in a quantum harmonic oscillator, this competition gives $w = \lambdabar=\xi_\text{M}=\sqrt{\hbar/(2m\Omega)}$. 

In Fig.~\ref{fig:6}{\ba}, we extract the domain-wall widths from both the $n_\pm$ images and the phase maps. In all cases, one-dimensional cuts perpendicular to the walls are fitted well by the theoretically predicted~\cite{Son:2002} forms $n_{\pm}(z) = n_0\pm n_1\text{sech}^2(z/w)$ and $\phi(z) = 4\arctan[\exp(z/w)]$; here, $z$ is the coordinate perpendicular to a wall centered at $z=0$, and $n_0$, $n_1$ and $w$ are fit parameters.
The extracted values of $w$ have a wide distribution (see histograms), likely because our domain walls are excited, but the three mean values are (within one standard deviation) consistent with each other and with the theoretically predicted $w$, indicated on the histograms by the dashed lines; note that here we include an $8\%$ correction to the predicted $w$ due to our nonzero $Z_0$ (Methods).

Finally, in Fig.~\ref{fig:6}\bb, we show that the characteristic wall width responds to changes in $\Omega$. Here, we initiate the system dynamics as in Fig.~\ref{fig:5}{\bc}, but quench $\Omega$ either down or up for the last $2\,$ms of relaxation. In the representative $n_+$ images one can see the systematic change in the wall widths, and the widths extracted as in Fig.~\ref{fig:6}{\ba} are consistent with the expected trend $w \propto 1/\sqrt{\Omega}$.}

To conclude, we have established a versatile { analog} simulator for self-interacting relativistic { quantum} fields with a tunable mass in $2+1$ dimensions, and observed non-perturbative phenomena that point the way to many future studies.
{  The damping of the global $\phi$ oscillations, accompanied by an emergence of spatial structures ($k>0$ excitations), is a process closely analogous to cosmological preheating and reheating~\cite{Kofman:1994, Chatrchyan:2021}.} 
The $\pi$ phase jump that places the field at the top of the sine-Gordon potential initiates dynamics associated with a quench through a discrete symmetry-breaking phase transition (with the energy minima changing from $\phi=\pi$ to $\phi=0$ and $2\pi$). Generalizing the $\pi$ phase jump to a ramp of $\Omega$ from a positive to a negative value, one can study the associated Kibble--Zurek mechanism~\cite{Suzuki:2025}, linked to the formation of topological defects such as cosmic strings and domain walls~\cite{Vilenkin:1985, Vilenkin:2000}.  Studies of domain-wall networks could provide insights into cosmological dark matter production in axion scenarios~\cite{Marsh:2016, Benabou:2025} and gravitational-wave signals in a broad family of particle-physics models~\cite{Saikawa:2017}. 
{  In this context, a particularly interesting possibility is to use vector light shifts~\cite{Grimm:2000} to deterministically imprint specific initial domain-wall configurations and study their subsequent dynamics, including their relaxation and decay~\cite{Ihara:2019, Gallemi:2019}.}
Finally, using Floquet engineering to stabilize the system at $\phi=\pi$ would enable studies of false-vacuum decay~\cite{Coleman:1977a,Callan:1977, Zenesini:2024, Cominotti:2025} in a relativistic, cosmologically relevant setting~\cite{Fialko:2015, Braden:2018, Billam:2019,Braden:2019, Jenkins:2024}.

{\bf Acknowledgments}\quad We thank Silke Weinfurtner, Ian Moss, Hiranya Peiris, and Andrew Pontzen for inspiring discussions, and Tanish Satoor for experimental assistance. Our work was supported by EPSRC [Grant No.~EP/Y01510X/1], ERC [UniFlat], and STFC [Grants No.~ST/T006056/1 and No.~ST/Y004469/1]. 
A.C.J. was supported by EPSRC [Grant No. EP/U536684/1], and by a KICC Gavin Boyle Fellowship.
{ G.C. was supported by an NSERC Postdoctoral Fellowship.}
J.H.T.\ acknowledges hospitality and support from Trinity College, Cambridge.
Z.H.\ acknowledges support from the Royal Society Wolfson Fellowship.

%


\clearpage
\setcounter{figure}{0} 
\renewcommand\thefigure{\arabic{figure}} 
\renewcommand{\figurename}[1]{Extended Data Fig.~}

\section{Methods}

\textbf{Experimental setup.}
Our experiments are performed in a uniform two-dimensional (2D) optical box trap formed by repulsive $532\,$nm light. The 2D confinement is provided by an accordion lattice with trapping frequency $\omega_z/(2\pi)\simeq 1\,$kHz, generated holographically~\cite{Zupancic:2016} with a digital micro-mirror device (DMD). The in-plane confinement is provided by a square box directly projected from another DMD through a high resolution microscope (N.A. = 0.7). The box size is approximately $50\,\upmu{\rm m} \times 50\,\upmu{\rm m}$ for Fig.~\ref{fig:1}$\bc$, and $30\,\upmu{\rm m} \times 30\,\upmu{\rm m}$ for the rest of the experiments.

The RF field coupling the two hyperfine states is derived from a vector signal generator (Keysight N5172B). Using I/Q modulation, we independently and simultaneously control the amplitude, phase, and frequency of the RF field; to prevent carrier feed-through from contaminating the signal, we use a $200\,$kHz baseband offset. This allows us to dynamically attenuate the RF power by $>60\,$dB, and realize $\Omega/(2\pi)\lesssim10~$Hz while retaining the ability to perform fast $\pi/2$ rotation at $\Omega/(2\pi)>10~$kHz. {  The calibration error for $\Omega$ is $\lesssim1\%$ (std).}

To stabilize the magnetic field $B$, we first regulate the current through our coils using a feedback loop with a bandwidth $>1\,$kHz and then cancel the residual $50\,$Hz mains oscillations using feedforward. This cancels most of the high frequency noise from the power supply. To counter slow thermal drifts, we periodically (approximately once per minute) calibrate the field using partial Landau--Zener transfers. Based on $Z_\text{err}$ from the calibration data, we estimate the residual field variations to be about $150~\upmu$G (r.m.s.).

\textbf{Interaction properties.}
{ 
The 2D interaction parameters are $g_{\sigma\sigma'} = (\hbar^2/m)\sqrt{8\pi}\, a_{\sigma\sigma'}/\ell_z$, where $a_{\sigma\sigma'}$ are the $s$-wave scattering lengths, which we calculate using coupled-channel calculations~\cite{molscat:2019, Tiemann:2020}, and $\ell_z=\sqrt{\hbar/(m\omega_z)}$~\cite{Hadzibabic:2011}.

\begin{table*}
\caption{Experimental parameters. 
Here, we give the typical values for the densities and the chemical potentials.
{  The statistical variations in $\mu_\text{d}$ and $\mu_\text{s}$,  due to experimental fluctuations in the atom number, are typically 5$\%$ (std) within each dataset, and the uncertainties in their calibrations are approximately 3$\%$ (std).}
}
\centering
\begin{tabular}{cccccccc}
\toprule
$B$ (G) &
$a_{\uparrow\uparrow}\,\left(a_0\right)$ &
$a_{\uparrow\downarrow}\,\left(a_0\right)$ &
$a_{\downarrow\downarrow}\,\left(a_0\right)$ &
$Z_0$ &
$n$ $\rm\left(\upmu\mathrm{m}^{-2}\right)$ &
$\mu_\mathrm{s}/h\,\left(\mathrm{Hz}\right)$ &
$\mu_\mathrm{d}/h\,\left(\mathrm{Hz}\right)$ \\
\midrule
 $57.3$ &  $32$ & $-53$ & $110$ & $0.31$ & $100$ & $830$ & $40$ \\
$58.1$ & $31$ & $-53$ & $220$ & $0.53$ & $70$ & $850$ & $110$ \\
\bottomrule
\end{tabular}
\label{tab:expt-params}
\end{table*}

The interaction energy density is  
\begin{equation}
    \mathcal{E}_\text{int} = \frac{1}{2}\left(g_{\uparrow\uparrow} n_\uparrow^2 +  2 g_{\uparrow\downarrow} n_\uparrow n_\downarrow + g_{\downarrow\downarrow} n_\downarrow^2 \right)\,,
\end{equation}
which can be re-written as
\begin{equation}
    \mathcal{E}_\text{int}  =\frac{1}{2}g_\text{d} n^2 + \frac{1}{2}(Z - Z_0)^2 \kappa n^2 \, ,
\end{equation}
where $g_\text{d} = (g_{\uparrow\uparrow} g_{\downarrow\downarrow} - g_{\uparrow\downarrow}^2)/(g_{\uparrow\uparrow}+g_{\downarrow\downarrow} - 2g_{\uparrow\downarrow})$, $\kappa=(g_{\uparrow\uparrow}+g_{\downarrow\downarrow} - 2g_{\uparrow\downarrow})/4$, and $Z_0=-\Delta/\kappa$, with $\Delta=(g_{\uparrow\uparrow} - g_{\downarrow\downarrow})/4$. 

For $\Omega=\delta = 0$, the density and spin chemical potentials are, respectively, $\mu_\text{d}=g_\text{d}n$ and $\mu_\text{s}=\kappa n$. For $\Omega,\delta\neq0$, these quantities still provide the relevant interaction energy scales.}

For $\Omega \gg \mu_{\rm s}/\hbar$ and $\delta=0$, the plasma frequencies for oscillations near $\phi=0$ and $\phi=\pi$ are $\omega_{{\rm p},0} = \sqrt{\Omega(\Omega+2\mu_\text{s}/\hbar)}$ and $\omega_{{\rm p},\pi} = \sqrt{\Omega(\Omega-2\mu_\text{s}/\hbar)}$. By measuring $\omega_{{\rm p},0}-\omega_{{\rm p},\pi}\simeq2\mu_{\rm s}/\hbar$, we calibrate $\mu_{\rm s}$ and hence $n$.
We summarize the interaction parameters relevant for our experiments in Table~\ref{tab:expt-params}.

\begin{figure}[h!]
    \centering
    \includegraphics[width=\columnwidth]{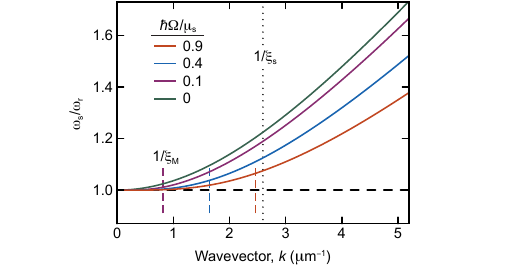}
    \caption{{  \textbf{Comparison of our spin dispersion $\omega_\text{s}(k)$ in Eq.~(\ref{eq:dispersion}) and the relativistic $\omega_\text{r}(k)$ in Eq.~(\ref{eq:relativistic dispersion}).} For $k < 1/\xi_{\rm s} \simeq 2.6~\upmu$m$^{-1}$ (dotted line), the difference between $\omega_\text{s}$ and $\omega_\text{r}$ is never larger than $\approx 20\%$. The colored dashed lines indicate $1/\xi_{\rm M}$ for the three nonzero $\hbar \Omega/\mu_{\rm s}$. For the domain walls in Figs.~\ref{fig:5} and \ref{fig:6}, $1/w$ is always smaller than the smallest $1/\xi_{\rm M}$ shown here, so they are deep in the relativistic regime, where the difference between our spin dispersion and the relativistic one is negligible. }}
    \label{fig:ExtendedFigure1}
\end{figure}

\textbf{Spin excitations.} 
{ 
Following Ref.~\cite{Jenkins:2024}, the Lagrangian, $\mathcal{L}$, that describes perturbative spin dynamics is
\begin{equation}
\begin{split}
    2\mathcal{L}/(\alpha n) = \: \hbar\,\delta\dot{\phi}~\, \delta Z' - \frac{1}{2}\delta\phi\left[\hbar \Omega' - \frac{\hbar^2\nabla^2}{2m}\right]& \delta\phi~\\
    \quad- \frac{1}{2}\delta Z'\left[(\hbar\Omega' + 2\mu_\text{s}') -\frac{\hbar^2\nabla^2}{2m}\right]& \delta Z'\,,
\end{split}
\label{eq:langrangian-full}
\end{equation}
Here, trigonometric corrections due to nonzero $Z_0$ are incorporated through the factor $\alpha = (1-Z_0^2)$ and the scaled variables $\Omega'= \Omega/\sqrt{\alpha}$, $\mu_\text{s}'=\alpha\mu_\text{s}$, and $\delta Z'=\, \delta Z/\alpha$.
From Eq.~(\ref{eq:langrangian-full}), we can derive various quantities such as the spin dispersion relation,

\begin{equation}
    \hbar \omega_\text{s}(k) =\sqrt{(\epsilon_k + \hbar\Omega')(\epsilon_k + \hbar\Omega' + 2\mu_\text{s}')} \, ,
    \label{eq:dispersion-full}
\end{equation}
and the aspect ratios for the plasma oscillations
\begin{equation}
\begin{split}
\begin{split}
    A_Z/A_\phi &= \alpha\sqrt{\hbar\Omega' / (\hbar\Omega' + 2\mu_\text{s}')}\, ,\\
    A_Z/A_Y & = A_Z/(\sqrt{\alpha} A_\phi)\, .
\end{split}
\end{split}
    \label{eq:aspect-ratio-full}
\end{equation}
}

{  
In the main text we give the simpler and more intuitive approximate equations that assume $Z_0 =0$. For all expressions in the main text other than Eqs.~(\ref{eq:langrangian}) and (\ref{eq:aspect-ratio}), for which the exact forms can be obtained from 
Eqs.~(\ref{eq:langrangian-full}) and (\ref{eq:aspect-ratio-full}),  the corrections due to nonzero $Z_0$ can be incorporated by simply replacing $\Omega$ by $\Omega'$ and $\mu_\text{s}$ by $\mu_\text{s}'$.
For our experiments, these corrections are usually negligible (compared to the statistical errors), as we illustrate in Fig.~\ref{fig:3}{\bc} by plotting both the predictions that take into account the $Z_0$ corrections (dotted lines) and the approximate predictions that assume $Z_0=0$ (solid lines). For Figs.~\ref{fig:4} and \ref{fig:5}, the corrections are similarly negligible and for simplicity we ignore them, using only the simpler theory given in the main text. The only perceptible correction, which we do include when comparing the data to theory, is in Fig.~\ref{fig:6}, where, for $Z_0=0.53$, the expected wall width $w =\sqrt{\hbar/(2m\Omega')}$ is $\approx 8\%$ smaller than $\sqrt{\hbar/(2m\Omega)}$.

{  In Extended Data Fig.~\ref{fig:ExtendedFigure1}, we also examine the range of $k$ values for which the dispersion of our spin modes, $\omega_{\rm s}(k)$ in Eq.~(\ref{eq:dispersion}),  mimics well the relativistic dispersion $\omega_{\rm r}(k)$ in Eq.~(\ref{eq:relativistic dispersion}). 
As long as $k < 1/\xi_{\rm s}$, the difference between $\omega_{\rm s}$ and $\omega_{\rm r}$ is $\lesssim 20\%$ even for $\Omega =0$, and it reduces with increasing $\Omega$. 

Most importantly, the domain walls we observe in Figs.~\ref{fig:5} and \ref{fig:6} are deep in the long-wavelength, relativistic regime. For these walls, the relevant $k$ modes are $k \lesssim 1/w$, and even for our largest $\Omega$ in Fig.~\ref{fig:6}, $1/w$ is still smaller than the smallest $1/\xi_{\rm M}$ (purple dashed line) shown in Extended Data Fig.~\ref{fig:ExtendedFigure1}.  In this $k$ range, $\omega_{\rm s}(k)$ and $\omega_{\rm r}(k)$ are essentially identical.}

\begin{figure}[h!]
    \centering
    \includegraphics[width=\columnwidth]{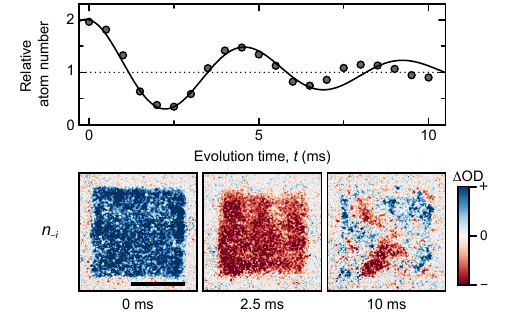}
    \caption{{ 
    \textbf{Damping of large-amplitude plasma oscillation.} Here, $Z_0=0.53$, $\Omega/(2\pi)\simeq45 \,$Hz, $\mu_{\rm s}/(\hbar\Omega) \simeq 18$, and we jump the phase by 0.5$\pi$. We image $n_{-i}$, which reveals the local $\sin[\phi(x,y)]$. In the top panel, spatially integrated $n_{-i}$ shows a damped plasma oscillation. The bottom images show that this damping is accompanied by emergence of spatial structures that correspond to a transfer of energy from the $k=0$ oscillation to higher-$k$ modes.  These long-time spatial structures vary randomly between experimental shots. The black scale bar shows $20\,\upmu$m.}}
    \label{fig:ExtendedFigure2}
\end{figure}

{ \textbf{Damping of plasma oscillations.} 
In Extended Data Fig.~\ref{fig:ExtendedFigure2} we illustrate how the large-amplitude plasma oscillations, such as shown in Fig.~\ref{fig:5}{\bb}, damp at long times. This damping is accompanied by emergence of spatial structures corresponding to a transfer of energy from $k=0$ to higher $k$ modes; qualitatively, this is what is expected in models of cosmological preheating and reheating~\cite{Kofman:1994, Chatrchyan:2021}.
}

\begin{figure}[b]
    \centering
    \includegraphics[width=\columnwidth]{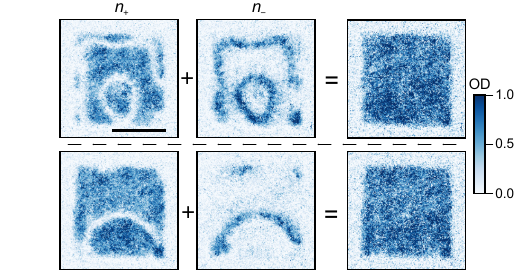}
    \caption{{ \textbf{Total density at the domain wall.} Here we show examples of $n_+$ and $n_-$ density distributions imaged in the same experimental run, with the protocol otherwise as for Fig.~\ref{fig:5}{\bc}. The two density distributions are complementary, and we do not observe any significant variation of the total density, $n_+ + n_-$. The black scale bar corresponds to $20\,\upmu$m. Here $Z_0=0.53$, $\Omega/(2\pi)\simeq22\,$Hz, and $\mu_{\rm s}/(\hbar\Omega) \simeq 37.$}} 
    \label{fig:ExtendedFigure3}
\end{figure}

{ \textbf{Total density at the domain wall.} In Extended Data Fig.~\ref{fig:ExtendedFigure3}, we create domain walls as in Fig.~\ref{fig:5}{\bc}, and take consecutive images of $n_{+}$ and $n_{-}$ in the same experimental run, which shows that the two density distributions are complementary.}

{ \textbf{Phase reconstruction.} 
From our $n_{\rm RB}(x,y)$ images, we extract $\phi_\text{tot} = \mathbf{q}\cdot\mathbf{r} + \phi$ using a convolutional neural network~\cite{Ronneberger2015}, trained on synthetic data that mimics the experimental images. The network generates quadrature maps $I(x, y)$ and $Q(x, y)$ that approximate, respectively, $\cos[\phi_\text{tot}(x, y)]$ and $\sin[\phi_\text{tot}(x, y)]$, and we extract $\phi_\text{tot}(x, y)$ as the phase of $I(x, y) + iQ(x, y)$. 

In Extended Data Fig.~\ref{fig:ExtendedFigure4}, we illustrate the fidelity of this extraction for two examples shown in Fig.~\ref{fig:5}{\bd} and two additional ones from the same dataset. Left to right, we show: (1) the $n_{\rm RB}$ images, (2) the reconstructed $ [1 + \cos(\phi_\text{tot})]/2$, which is essentially the de-noised version of the fringes seen in $n_{\rm RB}$, (3) the contours tracing the maxima of the reconstructed fringes, (4) the overlaying of these contours on the original images, and (5) the extracted $\phi(x,y)$.

We obtain $\phi(x,y)$ from $\phi_\text{tot}(x,y)$ by subtracting the independently calibrated carrier phase $\mathbf{q}\cdot \mathbf{r}$, but note that, because the relative phase between the RB beams and the RF coupling is random, our measurement determines $\phi(x,y)$ only up to a global offset. We fix this offset such that $\langle e^{i\phi}\rangle = -1$ for $t=0$\,ms (since $\phi$ is jumped to $\pi$) and $\langle e^{i\phi}\rangle = 1$ for $t=10$\,ms (appropriate for relaxation to  $\phi = 0\pmod{2\pi}$, as observed in the $n_\pm$ images).

}

\begin{figure*}
    \centering
    \includegraphics[width=\textwidth]
    {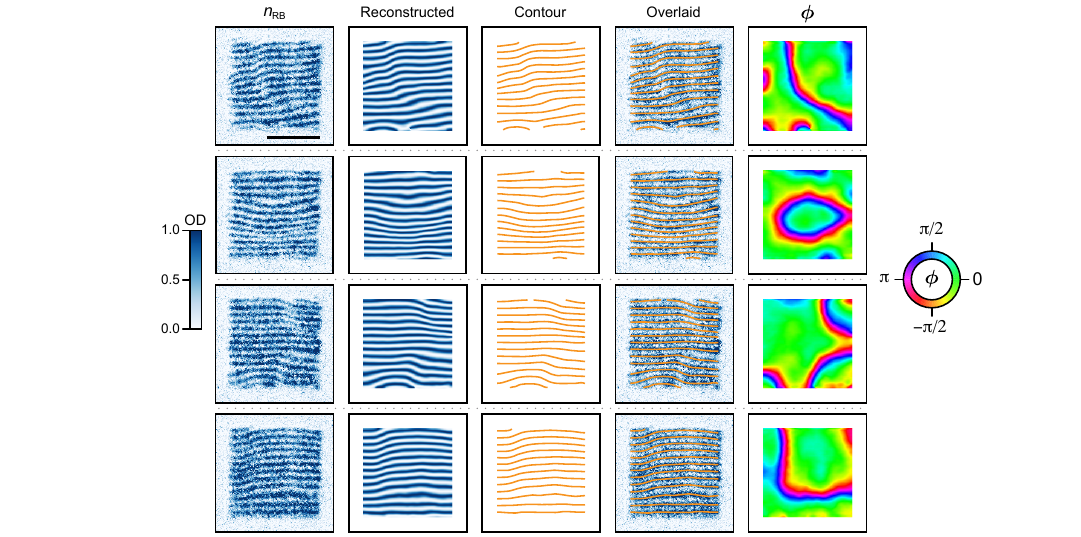}
    \caption{ 
    \textbf{Phase reconstruction.}
     We illustrate our phase-extraction process and its fidelity for two examples from Fig.~\ref{fig:5}\bd~and two additional ones. We extract $\phi_\text{tot}=\mathbf{q}\cdot \mathbf{r} + \phi$ from the $n_\text{RB}$ images using a convolutional neural network~\cite{Ronneberger2015} trained on synthetic data. Here we show the reconstructed fringe pattern $[ 1 + \cos(\phi_\text{tot})]/2$, the contours tracing the reconstructed-fringe maxima, and the overlays of these contours on the original experimental images. To obtain $\phi(x,y)$, we subtract from $\phi_\text{tot}$ the independently calibrated carrier phase $\mathbf{q}\cdot \mathbf{r}$.
    This determines $\phi$ only up to a global offset, because the relative phase between the RB beams and the RF coupling is random from shot to shot. We fix this offset such that $\langle e^{i\phi}\rangle$ matches the expected value. For the data in Fig.~\ref{fig:5}\bd, we impose $\langle e^{i\phi}\rangle = -1$ for $t = 0$\,ms and $\langle e^{i\phi}\rangle = 1$ for $t = 10$\,ms.  The scale bar shows $20\,\upmu$m and the fringe period $2\pi/q$ is $2.51\pm0.04\,\upmu$m, with the error originating from Fourier broadening.
    }
    \label{fig:ExtendedFigure4}
\end{figure*}


\end{document}